\documentclass{article}
\usepackage{spconf,amsmath,graphicx}

\copyrightnotice{978-1-6654-7189-3/22/\$31.00~\copyright2023 IEEE}

\usepackage[table]{xcolor}
\usepackage[T1]{fontenc}
\usepackage[hidelinks]{hyperref}
\usepackage{booktabs}
\usepackage{multirow}

\definecolor{Gray}{gray}{0.9}

\title{BERTraffic: BERT-based Joint Speaker~Role and Speaker~Change Detection for Air Traffic Control Communications}

\name{
\begin{tabular}{c}
Juan Zuluaga-Gomez$^{~\star, \dagger, \ddagger}$, 
Seyyed Saeed Sarfjoo$^{~\dagger}$,
Amrutha Prasad$^{~\dagger, \mathparagraph}$, 
Iuliia Nigmatulina$^{~\dagger}$, \\
Petr Motlicek$^{~\dagger,\mathparagraph}$, 
Karel Ondrej$^{~\mathparagraph}$,
Oliver Ohneiser$^{~\mathsection}$, 
Hartmut Helmke$^{~\mathsection}$
\end{tabular}
\thanks{Copyright 2023 IEEE. Published in the 2022 IEEE Spoken Language Technology Workshop (SLT) (SLT 2022), scheduled for 19-22 January 2023 in Doha, Qatar. Personal use of this material is permitted. However, permission to reprint/republish this material for advertising or promotional purposes or for creating new collective works for resale or redistribution to servers or lists, or to reuse any copyrighted component of this work in other works, must be obtained from the IEEE.}
\thanks{$^{\star}$Corresponding author: juan-pablo.zuluaga@idiap.ch}
}

\address{
  $^{\dagger}$ Idiap Research Institute, Martigny, Switzerland \\
  $^{\ddagger}$ Ecole Polytechnique Federale de Lausanne (EPFL), Switzerland \\
  $^{\mathparagraph}$ Brno University of Technology, Brno, Czech Republic \\
  $^{\mathsection}$ German Aerospace Center (DLR), Institute of Flight Guidance, Braunschweig, Germany \\
}

\begin{document}
\ninept
\maketitle
%
\begin{abstract}
Automatic speech recognition (ASR) allows transcribing the communications between air traffic controllers (ATCOs) and aircraft pilots. The transcriptions are used later to extract ATC named entities, e.g., aircraft callsigns. One common challenge is speech activity detection (SAD) and speaker diarization (SD). In the failure condition, two or more segments remain in the same recording, jeopardizing the overall performance. We propose a system that combines SAD and a BERT model to perform speaker change detection and speaker role detection (SRD) by chunking ASR transcripts, i.e., SD with a defined number of speakers together with SRD. The proposed model is evaluated on real-life public ATC databases. Our BERT SD model baseline reaches up to 10\% and 20\% token-based Jaccard error rate (JER) in public and private ATC databases. We also achieved relative improvements of 32\% and 7.7\% in JERs and SD error rate (DER), respectively, compared to VBx, a well-known SD system.\footnote{Our code is stored in the following public GitHub repository: \url{https://github.com/idiap/bert-text-diarization-atc}}
\end{abstract}


\section{Introduction}

Air traffic controllers (ATCOs) supervise a portion of airspace by issuing commands to pilots. Most of these voice-based communications are conveyed over noisy VHF (very-high frequency) channels, i.e., low signal-to-noise ratio (SNR). In a typical scenario, the ATCO (speaker1) issues voice-based commands to pilots (speaker2) together with pre-defined callsigns (names of the aircraft). Considering that a big portion of this communication is transmitted via voice messages, previous studies proposed to apply automatic speech recognition (ASR) to automatically extract the corresponding transcripts and high-level entities. 
In recent years, the ASR systems were shown to reach maturity in reducing ATCO's workload, but for research-only scenarios. Examples are AcListant{\textregistered}-Strips~\cite{helmke2017increasing} and MALORCA\footnote{\url{https://www.malorca-project.de}} projects. The later shows that novel data-driven machine learning approaches enable costly adaptations to different airport environments~\cite{kleinert2018semi}. Lin~\cite{lin2021spoken,yi2022identifying} reviews ten tasks on spoken instruction understanding of air traffic control (ATC) data. Semi-supervised learning has also been explored on the framework of ATC~\cite{srinivasamurthy2017semi}. HAAWAII\footnote{\url{https://www.haawaii.de}} and SOL-Cnt projects focus on developing a reliable and adaptable ASR engine for transcribing ATCO-pilot ATC communications. Previous work has concluded that higher accent variability and noise level cause ASR systems to yield up to two times higher word error rates (WER) for pilots' utterances compared to ATCOs' utterances~\cite{pellegrini2018airbus}. 
In addition, close and cross-talk between ATCO and pilots induce acoustic-based speaker diarization (SD) systems to yield non-acceptable performances. All this together jeopardizes the speaker change detection (SCD) step and subsequently the ASR system ends up processing utterances with multiple speakers. 

\begin{table}[t]
  \caption{Conversation between two speakers with correct SAD and SCD (rows 1 and 2) and SCD fault (row 3, words in bold). $^\dagger$samples from \textit{SOL-Cnt} test~set.}
  \vspace{0.1cm}
  \label{tab:improve_examples}
  \centering
  \begin{tabular}{ p{2.3cm}p{4.7cm} }
    \toprule
    Speaker Label & Detected segment$^\dagger$ \\
    \midrule
    ATCO (speaker 1) & \textcolor{blue}{<s> november six two nine charlie tango report when established </s>} \\
    \midrule
    Pilot (speaker 2) & \textcolor{magenta}{<s> report when established november six two nine charlie tango </s>} \\
    \midrule
    \raggedright Mixed (SAD and SCD failed) & \textcolor{blue}{<s> november six two nine charlie tango report when established} \textbf{report} 
    \textcolor{black}{\textbf{when established} </s>} \textcolor{magenta}{<s> november six two nine charlie tango </s>} \\
    \bottomrule
  \end{tabular}
\end{table}

\subsection{Motivation}

Already existent acoustic-based SD systems, like~\cite{landini2022bayesian} or end-to-end neural-based SD~\cite{fujita2019end_free}, show promising performances for many applications. However, in ATC communications, given its limitations such as high speaker rate, close-talk, and noise levels, relying solely on the acoustic level has shown to be insufficient. Additionally, standard SD systems add one layer of complexity to the whole ATC pipeline,\footnote{A standard ATC pipeline is composed of signal processing, SAD and SD, ASR, natural language understanding and post-processing.} weakening the flexibility to transfer the already tuned pipelines to other environments (e.g., noise level variation or new speakers' accents). That is why applying SD solely on the text level stands as an interesting solution to target these disadvantages. 
Additionally, the proposed SD system is speaker-agnostic because it fed with text data. This, can drastically reduce the chance of speaker identification as we remove the possibility to obtain the speaker identity from acoustic data or features. 

\subsection{Contribution}

\begin{figure*}[h]
  \centering
  \includegraphics[width=0.9\textwidth]{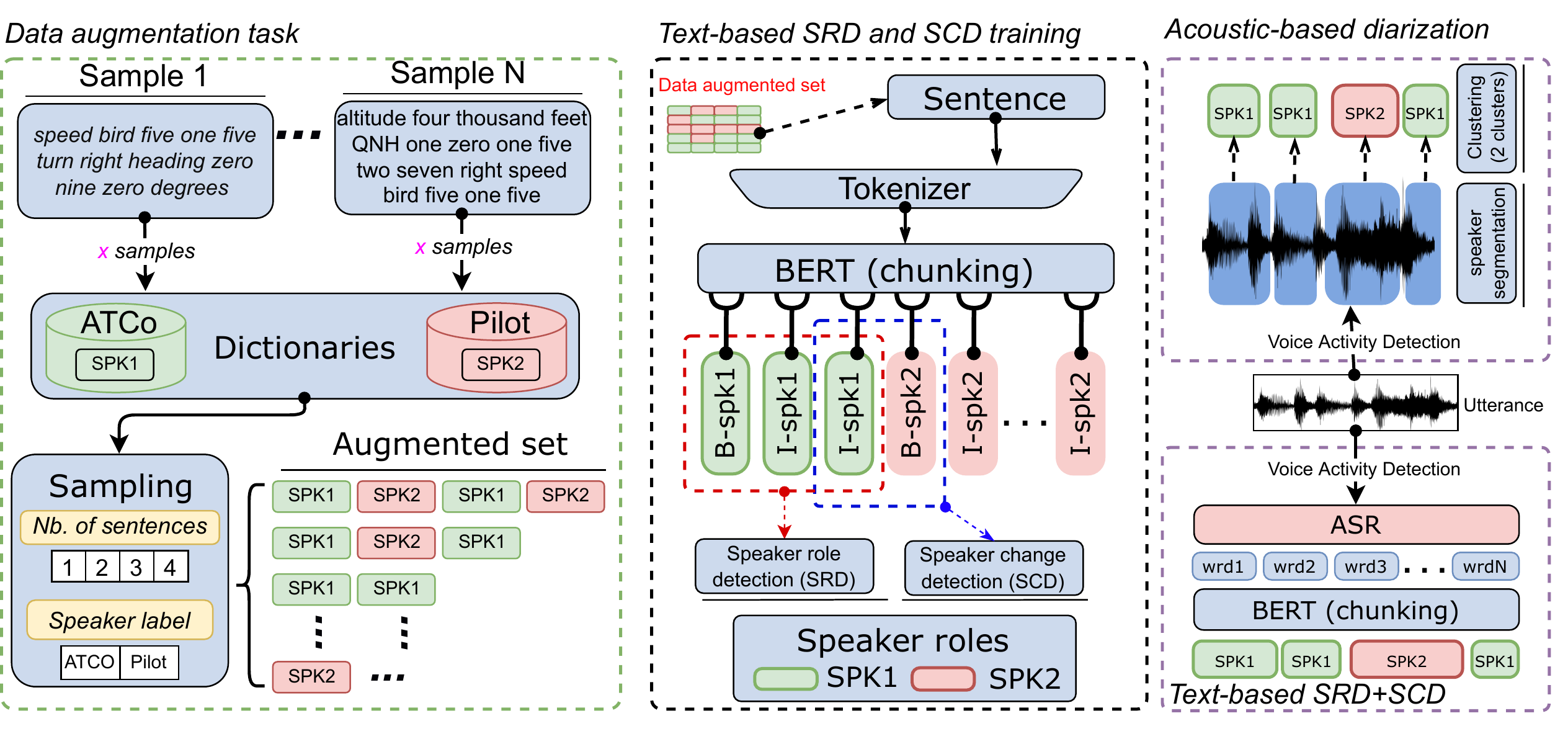}
  \caption{\textbf{Left block:} proposed data augmentation pipeline. Augmented samples contain between one and four utterances (probabilities of 40\%, 30\%, 20\% and 10\% for one to four, respectively). New sentences have equal chance to be sampled from the ATCO or pilot dictionary. \textbf{Central block:} proposed pipeline to fine-tune a BERT model that performs tagging and chunking for joint SCD and SRD. \textbf{Right block:} proposed approach to compare acoustic-based SD (VBx) and BERT joint SRD and SCD. 
  }
  \label{fig:diarization_pipeline}
  \vspace{-4mm}
\end{figure*}

In this work, we fine-tune a pre-trained BERT model~\cite{devlin2018bert} to jointly perform tagging and chunking for SCD and speaker role detection (SRD). Chunking allows splitting sentences into tokens (or words) and then merging them in meaningful subgroups. In our case, a phrase (or entity) is composed of a full single-speaker utterance, where either ATCO or pilot is the role (see Table~\ref{tab:improve_examples}). By applying chunking in a multi-speaker and multi-segment (or single-speaker and single-segment) utterance, one can perform speaker change detection (SCD) and speaker role detection (SRD) simultaneously on the text level (Figure~\ref{fig:diarization_pipeline} mid-box). 
In short, our approach simplifies the standard SD pipeline, moving up the task from the acoustic level to text level, i.e., post ASR. We stack the BERT model on top of a speech activity detection (SAD) module to create a text-based SD, which from now on we call \textit{\textbf{`BERT SD system'}}. 
Our approach is proved on public and private databases. We developed a simple yet effective data augmentation pipeline to counteract the class imbalance within the train sets.
The BERT SD system (i.e, combination of SAD, SRD, and SCD) yielded acceptable token-based JER of about 10\% for seen domains (i.e., text transcripts provided to fine-tune the BERT model) and no more than 20\% JER for databases that have not seen during training (in this case, the model has been fine-tuned on the SD task with other in-domain text data). Finally, we also experimented by directly feeding the BERT-based SD with transcripts generated by our in-domain hybrid-based ASR system for ATC~\cite{zuluagagomez21_interspeech,kocour2021automatic}. We obtained competitive performances compared to acoustic-based SD baselines. 



\section{Related Work}

Speaker diarization systems answer the question \textit{``who spoke when?''}. SAD, segmentation or SCD, embedding extraction, clustering and labeling are the main parts of a SD system. 

\textbf{Traditional acoustic-based diarization}: feature representations of speakers are one of the main factors in the accuracy of a SD system. Mel frequency cepstral coefficients (MFCCs) are commonly used for the task of SD. In comparison to MFCC, mel filterbank slope (MFS) and linear filterbank slope (LFS) features have more speaker discriminability power caused by emphasis on higher-order formants~\cite{madikeri2014filterbank}. The agglomerative information bottleneck (aIB) approach to SD has shown competitive performance~\cite{vijayasenan2009information}. Here, for clustering the fixed-length audio segments, a bottom-up clustering approach is applied in the posterior space represented by a mixture of Gaussians. Speaker discriminative embeddings such as x-vectors are investigated in~\cite{sell2018diarization}. For finding the speaker clusters in a sequence of x-vectors, the variational Bayesian hidden Markov model (VBx) was investigated in~\cite{valente2010variational,landini2022bayesian}. For continuously learning speaker discriminative information, ``Remember-Learn-Transfer'' was proposed in~\cite{dawalatabad2019incremental}. 
Applying lexical and acoustic information for SD was investigated in~\cite{park2018multimodal}.


\textbf{Neural-based diarization}: in the last years, there has been an increasing interest in end-to-end (E2E) and sequence-to-sequence architectures for different speech-related tasks. SD and its derivates, e.g., SCD, have also seen the benefits from this trend. For example,~\cite{el2019joint} builds on top of their proposed baseline for SD (ASR and SD are run in parallel and then the output is merged). They perform joint ASR and SD, claiming that word-level DER can be improved up to 15.8\% in cross-domain evaluations. Afterward, the end-to-end neural diarization (EEND) was introduced in~\cite{fujita2019end_free}, where a full SD model is trained jointly to perform extraction and clustering. Later, the same authors upgraded the system by replacing the bidirectional long short-term memory (BLSTM) layers by self-attention modules~\cite{fujita2019end}. Subsequent work has targeted EEND for unknown number of speakers~\cite{horiguchi2020end}, SD for long conversations~\cite{mao2020speech}, streaming EEND~\cite{han2021bw}, SD constrained by turn detection (i.e., SCD) in~\cite{xia2022turn}, or even leveraging EEND for ASR~\cite{khare2022asr}.

\textbf{Text-based speaker role detection}: early text-based techniques for SRD or SCD relied on handcrafted lexicons, dictionaries, and rules. They are prone to human errors and not robust against noisy labels, e.g., produced by standard ASR systems (e.g.,~\cite{valente2011speaker}).
Collobert et al.~\cite{collobert2011natural} introduced machine learning methods for text processing in part-of-speech tagging, chunking, and semantic role labeling. 
In~\cite{mohapatra10domain}, domain-based chunking of sentences is addressed, which is similar to the approach proposed in this paper. 
In general, chunking is used to parse phrases from unstructured text. In our case, tagging and chunking an ATC utterance allows us to perform jointly SCD and SRD.
The reader might relate chunking to named entity recognition (NER). NER is a chunking sub-task that aims at identifying entities on text, e.g., locations, organizations, or names~\cite{piskorski2017first,yadav2018survey,sharma2022named}. Examples of named entities in ATC communications are \textit{callsigns}, \textit{command types}, etc. These entities carry rich information that gives cues about the speaker's role (ATCO or pilot). A recent work aligned to ATC domain is reviewed in~\cite{prasad2021grammar}. Here, a grammar-based approach performs SRD on single-speaker utterances. In~\cite{ma-etal-2017-text} a text-based SRD for multiparty dialogues is proposed, but limited to SRD. Finally, text-based diarization has been proposed in the past by~\cite{han2021bw,khare2022asr}. However, these previous works do not take into account the text structure, grammar, and syntax. 

\textbf{Contrasting with previous work:} different to other systems, e.g., EEND or traditional acoustic-based SD, our model is fed directly with text data (for instance, transcripts). The field of ATC holds some limitations and advantages regarding SD, where already existent acoustic-based EEND systems could fail. Some limitations are: ATC audio is noisy (below 15 dB SNR) with close and cross-talk speech. Some advantages are: the number of speaker roles are known (in our case two, ATCO and pilot) and the grammar between the two speaker roles slightly differs. Our main idea is to leverage those advantages in order to show that a fully text-based joint SCD and SRD system can perform on par or even better than traditional acoustic-based SD. 
As a clarification, similar scenarios where our approach can be applied are call-centers or patient–physician conversations, where the number of speaker roles are defined beforehand and their grammar structure also differs. 

To summarize, the main difference between our BERT SD system and EEND roots on the fact that we use a standard BERT model~\cite{devlin2018bert} fine-tuned to ATC text data instead of crafting a SD neural network system. BERT\footnote{We use \emph{BERT-base-uncased} (110M params) for all the experiments.} is known for its ease and powerful ability to be fine-tuned on many tasks and corpus with minimum effort (e.g., amount of labeled data). It also performs well in low-resource scenario, which is the case in ATC.
Finally, as our system removes the `acoustic level' complexity and moves it to the text level, we demonstrate that mapping to the target domain when we have specific speaker roles is more efficient in the text level. For instance, data augmentation on text is simpler than on the acoustic level or one can modify easily the training data to adapt it to another scenario by merely altering the text.

\begin{table}[t]
    \caption{Amount of train and test data (\#\,train utterances / \#\,test utterances) for each class. \textit{ATCO} and \textit{pilot} columns are single-speaker samples, while \textit{Mixed} are utterances with two or more segments. $^\dagger$real-life ATC set where speech activity detection failed.}
    \vspace{0.1cm}  
    \label{tab:test_sets}
    \centering
    \resizebox{0.99\linewidth}{!}{
    \begin{tabular}{ lcccc }
        \toprule
        \rule{0pt}{3ex} \textbf{Database} & \multicolumn{1}{c}{ATCO} & \multicolumn{1}{c}{Pilot} & Mixed & Ref\\
        \midrule
        \rowcolor{Gray}
        \multicolumn{5}{c}{\textbf{\textit{Private databases}}} \\
        \midrule
        {\footnotesize\textbf{SOL-Cnt$^\dagger$}} & 662\,/\,138 & 945\,/\,204 & 535\,/\,205 & ~\cite{ohneiser21_interspeech} \\
        {\footnotesize\textbf{HAAWAII}} & 18724\,/\,1954  & 21099\,/\,2299 & - / - & ~\cite{zuluaga2022does} \\
        \midrule
        \rowcolor{Gray}
        \multicolumn{5}{c}{\textbf{\textit{Public databases}}} \\
        \midrule
        {\footnotesize\textbf{ATCO2}} & - /\,1772 &  - /\,1350  & - / - & \cite{kocour2021automatic}\\
        {\footnotesize\textbf{LDC-ATCC}} & 12694\,/\,1515 &  14216\,/\,1446  & - / - & \cite{LDC_ATCC} \\
        {\footnotesize\textbf{UWB-ATCC}$^\dagger$} & 4577\,/\,1157 &  6669\,/\,1713  & 735/174 & \cite{UWB_ATCC} \\
        \bottomrule
    \end{tabular}
    }
\end{table}

\section{Databases and Experimental Setup}
\label{sec:datasets}


This research experiments on five databases in the English language with various accents and data quality. With the aim of encouraging open research on ATC (which has lagged behind due to privacy clauses and contracts\footnote{Nearly all ongoing and former projects in the area of ATC prohibit the release of code and models due to privacy issues.}) we identified and experimented with three public databases, as referenced in Table~\ref{tab:test_sets}. All experiments use 10\% of the train set as validation set. We release a GitHub repository
\footnote{\scriptsize\url{https://github.com/idiap/bert-text-diarization-atc}}
with training scripts to replicate the results on the public databases. To the author's knowledge, this is the first open release of code in the field of natural language processing for air traffic control. 

\subsection{Private databases}
\label{subsec:datasets}

\quad\quad \textbf{SOL-Cnt}: private database recorded and collected over EU-funded industrial research project that aims to reduce ATCOs' workload with an ASR-supported aircraft radar label. Voice utterances of ATCOs and pilots have been recorded in the operations' room at the air navigation service provider (ANSP) site of Austrocontrol in Vienna, Austria\footnote{PJ.16-W1-04:~\url{https://www.sesarju.eu/projects/cwphmi}}~\cite{ohneiser21_interspeech}.

\textbf{HAAWAII}: private data set that has been collected and annotated from ATC communications from London and Icelandic airspace.\footnote{\url{https://www.haawaii.de}} This data is of higher quality ($\geq$ 15 dB SNR) compared to SOL-Cnt. All the data is correctly split, i.e., one speaker per segment. Previous benchmark and results are covered in~\cite{nigmatulina2022two,zuluaga2022does}.

\subsection{Public databases}

\quad\quad \textbf{LDC-ATCC}: public ATC corpus gathered from three different airports in the US.
\textbf{LDC-ATCC}{\footnote{\url{https://catalog.ldc.upenn.edu/LDC94S14A}}} comprises recorded speech with the aim of supporting research in robust ASR. The recordings contain several speakers, and they were collected over noisy channels. The database is formatted in NIST Sphere format, where full transcripts, start and end times of each transmission are provided~\cite{LDC_ATCC}.

\textbf{UWB-ATCC}: public ATC corpus containing recordings of communication between ATCOs and pilots. The speech is manually transcribed and labeled with the information about the speaker (pilot/controller, not the full identity of the person). The audio data is single channel sampled at 8kHz. Similar to SOL-Cnt, UWB-ATCC contains around 900 utterances where segmentation failed and two or more segments and/or speakers ended up in the same recording. This database can be downloaded for free in their website
{\footnote{\url{https://lindat.mff.cuni.cz/repository/xmlui/handle/11858/00-097C-0000-0001-CCA1-0}}}~\cite{UWB_ATCC}.

\textbf{ATCO2 corpus}: public ATC corpus recently released in ATCO2 project.{\footnote{ATCO2 corpus: \url{https://www.atco2.org/data}}} ATCO2 developed a pipeline to pseudo-annotate (ASR transcripts, language and diarization labels) large amounts of ATC speech audio for training robust ASR models. We use the entire database only as test set (over 4000 utterances), thus we consider this as an out-of-domain evaluation.
The ATCO2 corpus is one of the few open-source and public databases that has been used by other researchers to benchmark their ASR engines~\cite{kocour2021automatic,zuluaga2020callsign}. The full corpus is available for purchase through ELDA in \url{http://www.elra.info/en/catalogues/}.

\subsection{Annotation protocol}
\label{sec:anno-protocol}

In addition to manual speech transcripts, speaker labels and time segmentation (e.g.,  ATCO/pilot/mixed) are also available. The BERT model starts by tagging each word of the transcript (ground truth or ASR transcript) with a set of tags that follows the well-known \textit{IOB format} (Inside-Outside-Beginning). In IOB format, each entity (a full sentence in our case) is composed of two tags: (i) the \textit{Beginning} defines which token/word is the start of the sentence \textit{\textbf{`B-'}}, and (ii) the \textit{Inside} tag \textit{\textbf{`I-'}} defines which tokens/words belongs to that specific sentence. We define ATCO recordings as \textit{Speaker1}, while pilot segments as \textit{Speaker2} (green and red in Figure~\ref{fig:diarization_pipeline}, respectively). We do not use the \textit{Outside} tag, because we know that each word is always from one of two predefined speakers. In total, we have four tags, two per class (ATCO and pilot).

\subsection{Data augmentation}
\label{sec:data-augmentation}

We implemented a simple yet effective data augmentation pipeline to counteract the class imbalance in the train sets (see Table~\ref{tab:test_sets}). First, we split the training sets on either \textit{ATCO} (speaker 1) or pilot (speaker 2) subset. Then, we generate new sentences from the initial set of utterances for each database (e.g., HAAWAII $\sim$39k utterances). Each new sample depends on: (i) the number of sentences to be concatenated, and (ii) the speaker label of each sentence. In general, a new sample is composed of one to four sentences, each with an equal chance of being drawn from the ATCO or pilot dictionary. The process is repeated until gathering $\sim$350\,MB of text data ($\sim$1M sentences). 
We emphasize that in ATC, there is no need to have a correlation between previous and next sentences/utterances. This is due to the fact that speaker 1 (ATCO) communicates to several speakers 2 (pilots). The stream of information received and transmitted by the speakers is not dependent on `left' or `right' context. Therefore, concatenating various segments randomly would not degrade substantially the WERs.\footnote{We measure WERs by decoding with the in-domain ASR system the original and augmented test sets to corroborate this assumption.
The relative WER degradation was less than 1\% for all test sets.} The left block in Figure~\ref{fig:diarization_pipeline} depicts the proposed data augmentation pipeline.

\begin{table*}[t]
    \caption{Jaccard error rate (JER) in percentages (\%) from predictions using different train (column 1) and test sets. All the experiments use the same model (BERT-base-uncased) and same set of hyperparameters. We report the mean of five runs with different seeds and its standard deviations (mean $\pm$ STD). \textbf{Bold} refers to the best performance over public databases, while \underline{underline} denotes the highest performance per column. Metrics reported on token level of ground truth transcripts.}
    \vspace{0.05cm}  
    \label{tab:overall_performance}
    \centering
    \begin{tabular}{lr|lcccccc}
        \toprule
        \multicolumn{2}{c}{\textbf{Model}} &  & \multicolumn{3}{c}{\cellcolor{Gray}\textbf{Public}} &  & \multicolumn{2}{c}{\cellcolor{Gray} \textbf{Private}} \\
        \cline{4-6} \cline{8-9}
        
        \rule{0pt}{3ex} \textbf{Database} & \textbf{\# samples} &  & \textbf{ATCO2} & \textbf{UWB-ATCC} & \textbf{LDC-ATCC} &  & \textbf{HAAWAII} & \textbf{SOL-Cnt} \\
        
        \midrule
        \multicolumn{9}{c}{\cellcolor{Gray} \textbf{Public databases}} \\
        \midrule
        
        LDC-ATCC & 26.9k &  & 31.3 $\pm$ 2.4 & 35.8 $\pm$ 2.0 & 8.1 $\pm$ 0.7 &  & 28.7 $\pm$ 3.1 & 52.6 $\pm$ 1.3 \\
        UWB-ATCC & 11.2k &  & 21.6 $\pm$ 0.7 & \underline{\textbf{10.7 $\pm$ 0.6}} & 18.7 $\pm$ 2.6 &  & 15.2 $\pm$ 1.4 & \underline{\textbf{18.7 $\pm$ 1.7}} \\
        $\hookrightarrow$ + LDC-ATCC & 38.1k &  & \textbf{19.8 $\pm$ 0.9} & 11.3 $\pm$ 0.4 & \underline{\textbf{7.1 $\pm$ 1.3}} &  & \textbf{14.2 $\pm$ 1.4} & 24.0 $\pm$ 1.9 \\
        
        \midrule
        \multicolumn{9}{c}{\cellcolor{Gray} \textbf{Private database}} \\
        \midrule
        
        HAAWAII & 39.8k &  & 23.9 $\pm$ 0.6 & 22.3 $\pm$ 1.7 & 14.1 $\pm$ 1.2 &  & 6.5 $\pm$ 0.7 & 48.5 $\pm$ 1.4 \\
        $\hookrightarrow$ +LDC+UWB & 77.9k &  & \underline{17.5 $\pm$ 0.2} & 11.5 $\pm$ 0.5 & 7.5 $\pm$ 0.6 &  & \underline{6.2 $\pm$ 0.3} & 26.8 $\pm$ 2.0 \\
        \bottomrule
    \end{tabular}
\end{table*}

\subsection{Modules}
\label{sec:modules}

The performance of our BERT-based SRD and SCD system is contrasted with a standard acoustic-based SD system. We use an out-of-the-box VBx system to evaluate the \textit{SOL-Cnt} and \textit{UWB-ATCC} test sets, which contain real-life ATC audio where segmentation failed. 
For both, BERT and acoustic-based SD systems, we use the same multilingual ASR-based SAD module~\cite{sarfjoo_interspeech2021} to remove the silence in the recording files.
\vspace{0.1cm}

\textbf{Speaker role and speaker change detection module:} the SRD and SCD systems are built on top of a pre-trained BERT model~\cite{devlin2018bert} downloaded from HuggingFace~\cite{wolf-etal-2020-transformers,lhoest-etal-2021-datasets}. The model is later fine-tuned with either the original or the augmented databases, on the tagging and chunking task (following~\textit{IOB}~format). 
We append a linear layer with a dimension of 4 (following the classes structure from Section \ref{sec:anno-protocol}) on top of the last layer of the BERT model. Then, we fine-tune each model on an NVIDIA GeForce RTX 3090 for 3k steps, with a learning rate scheduler that first warms up the learning rate until $\gamma=5\mathrm{e}{-5}$ for 500 steps, and then it linearly decays. 
We employ AdamW~\cite{loshchilov2018decoupled} optimizer ($\beta_1{=}0.9, \beta_2{=}0.999$, $\epsilon{=}1\mathrm{e}{-8}$) and dropout~\cite{srivastava2014dropout} of $dp=0.1$ for the attention and hidden layers. We use GELU activation function~\cite{hendrycks2016gaussian}. 
We train all models with batch size of 32, and gradient accumulation of 2, i.e., effective batch size of 64.

\vspace{0.1cm}

\textbf{Acoustic-based diarization:} 
for details of the VBx model, the reader is referred to~\cite{landini2022bayesian}. This model uses a Bayesian hidden Markov model (BHMM) to find speaker clusters in a sequence of x-vectors. Here, the x-vector extractor uses DNN architecture based on ResNet101. The input to the ResNet is 64 log Mel filter bank features extracted every 10 msec using 25 msec window. In the first step, Agglomerative Hierarchical Clustering (AHC) is applied to the extracted x-vectors. Then, Variational Bayes HMM over x-vectors is applied using the AHC output. For achieving the best performance on the database with short duration files with a maximum of two speakers, we tuned the probability of not switching speakers between frames (loopP) and speaker regularization coefficient (Fb) to 0.7 and 6, respectively. 
\vspace{0.1cm}

\textbf{Automatic speech recognition:} a state-of-art hybrid-based ASR system for ATC speech was developed with Kaldi toolkit~\cite{povey2011kaldi}. The system follows the standard recipe, e.g., uses MFCC and i-vectors features with standard chain training based on lattice-free MMI. We use the same ASR system for audio from both speakers (i.e., ATCO and pilot). The training recipe and databases (including the train sets in Table~\ref{tab:test_sets}) are covered in~\cite{kocour2021automatic,zuluagagomez20_interspeech,nigmatulina2021improving,nigmatulina2022two}.

\subsection{Evaluation protocol}

The experiments are prepared to answer three questions: (i) how reliable is the BERT-based SRD and SCD system on ground truth transcripts? (ii) How is the performance impacted when using automatically generated (ASR) transcripts instead of ground truth transcripts?\footnote{This  is a real-life scenario where ASR transcripts are fed to the BERT SD system instead of ground truth transcripts.} And, (iii) which system performs better on real-life ATC speech data, text or acoustic-based SD?
\vspace{0.1cm}

\textbf{Acoustic-based diarization:} to score acoustic-based diarization, we use DER and Jaccard Error Rate (JER) as metrics. DER measures the fraction of time that the segment is not attributed correctly to a speaker or to non-speech which is defined in Equation~\ref{eq:der}:

\begin{equation}
\label{eq:der}
DER=\frac{\text{false alarm}+\text{miss detection}+\text{speaker confusion}}{\text{Total duration of speech in the reference file}},
\end{equation}

\noindent where false alarm is the duration of non-speech incorrectly classified as speech, missed detection is the duration of speech incorrectly classified as non-speech, confusion is the duration of speaker confusion, and total is the total duration of speech in the reference. JER is a recently proposed metric~\cite{ryant19_interspeech} that avoids the bias towards the dominant speaker, i.e., evaluating equally all speakers. The JER is defined in Equation~\ref{eq:jer}:

\begin{equation}
\label{eq:jer}
JER=1-\frac{1}{\text{\#speakers}}\sum_{\text{speaker}}\text{max}_{\text{cluster}}\frac{|\text{speaker}\cap\text{cluster}|}{|\text{speaker}\cup\text{cluster}|},
\end{equation}

\noindent where $\text{speaker}$ is the selected speaker from reference and \textit{$\text{max}_{\text{cluster}}$} is the cluster from the system with maximum overlap duration with the currently selected speaker.
\vspace{0.1cm}

\textbf{Speaker role detection:} we evaluate SRD with JER on the token level (which is more aligned to SD) on the five proposed test sets. To clarify, \textbf{SOL-Cnt} and \textbf{UWB-ATCC} databases contain utterances with more than one speaker per utterance. Thus, we test the SD capabilities of the proposed BERT-based system on these two test sets. 
Results are shortlisted in Table~\ref{tab:der_res}. We first analyzed the model performance on the ideal case, i.e., we used the ground truth audio annotations to obtain JERs per test set, thus assuming we have access to a perfect ASR system (0\% WER). These results are listed in Table~\ref{tab:overall_performance}. We employ the Scikit-learn\footnote{We use weighted Jaccard error rate score. It calculates metrics for each class (i.e., ATCO and pilot), and finds their average weighted by support (the number of true instances for each class). This accounts for label imbalance.} Python library to calculate these scores.
\vspace{0.1cm}

\textbf{Speaker change detection:} in addition to SRD, the BERT system performs SCD, i.e., central block in Figure~\ref{fig:diarization_pipeline}. 
We evaluated this task with DER and JER on one private (\textbf{SOL-Cnt}) and one public (\textbf{UWB-ATCC}) test set, which contains utterances with one or two speakers. The \textit{MIXED} column in Table~\ref{tab:der_res} list the results corresponding to SCD only on the multi-speaker segments. For creating the segments from the BERT-based SCD system, we used forced alignment between audio and ground truth text using the trained ASR module. This module is explained in Section~\ref{sec:modules}. Similarly, time information from the ASR output transcripts was used to create the segments of the BERT-based SD system on the ASR transcripts.

\begin{table*}[t]
  \caption{Comparison of acoustic-based VBx SD and text-based SD on \textit{ATCO}, \textit{PILOT}, and \textit{MIXED} subsets of SOL-Cnt, and UWB-ATCC test sets. \textbf{Bold} refers to the best performance. $^{\dagger}$the performance of acoustic diarization system. $^{\dagger\dagger}$proposed BERT model trained on all the available data with data augmentation and evaluated on ground truth annotations (\_GT) or ASR transcripts (\_ASR).}
  \vspace{0.05cm}  
  \label{tab:der_res}
  \centering
  \begin{tabular}{ lccccccc}
    \toprule
    \rule{0pt}{1ex} & \multicolumn{3}{c}{\cellcolor{Gray}\textbf{Sol-Cnt test set}} & &  \multicolumn{3}{c}{\cellcolor{Gray}\textbf{UWB-ATCC test set}} \\
    \rule{0pt}{1ex} & \multicolumn{1}{c}{\textbf{DER (\%) }$\downarrow$} & &  \multicolumn{1}{c}{ \textbf{JER (\%) }$\downarrow$} & & \multicolumn{1}{c}{\textbf{DER (\%) }$\downarrow$} & & \multicolumn{1}{c}{\textbf{JER (\%) }$\downarrow$} \\
    \cline{2-2} \cline{4-4} \cline{6-6} \cline {8-8}
    \rule{0pt}{3ex} \textbf{Model} & AT / PI / MIX & & AT / PI / MIX & & AT / PI / MIX & & AT / PI / MIX\\
    \midrule
    \rowcolor{Gray} \multicolumn{8}{c}{\textbf{Acoustic-based speaker diarization}} \\
    \textit{Acoustic\_VBx}$^{\dagger}$ & 5.8 / 7.8 / 10.3 & & 7.0 / 10.9 / 22.2 & & \textbf{0.8} / \textbf{1.2} / 14.4 & & \textbf{0.6} / \textbf{0.7} / 39.4 \\
    \midrule
    \rowcolor{Gray} \multicolumn{8}{c}{\textbf{Text-based speaker diarization}} \\
    \textit{BERT\_GT}$^{\dagger\dagger}$ & \textbf{2.4} / \textbf{2.4} / \textbf{8.9} & & \textbf{1.0} / \textbf{2.2} / \textbf{15.0} & & 1.2 / 1.7 / \textbf{5.8} & & 1.1 / 1.1 / \textbf{16.6} \\
    \textit{BERT\_ASR}$^{\dagger\dagger}$ & 3.0 / 3.7 / 9.5 & & 1.5 / 3.2 / 15.1 & & 1.6 / 1.5 / 6.9 & & 1.2 / 1.2 / 20.1 \\
    \bottomrule
  \end{tabular}
\end{table*}

\section{Results and Discussion}

\noindent \textbf{Baseline performance of BERT SD:} we discuss the results listed in Table~\ref{tab:overall_performance}. Here, we aim at evaluating two aspects of the BERT SD system. First, we assess how well the model behaves on out-of-domain corpora. We fine-tune BERT models on each database and evaluate it on all five test sets. We call this: \textit{transferability} between corpora. Second, we establish baselines on both, public and private databases. Each model is fine-tuned five times with different seeds, hence we report the mean and standard deviation across runs. Not to our surprise, test data that matched the fine-tuning one performed particularly well. LDC-ATCC and UWB-ATCC test sets reached less than 10\% JER, while $\sim$20\% JER for ATCO2. 

One aspect that can shed light on new research is how public databases transfer to private ones. This can help future research to set a starting point, thus reducing the costs inherit by developing tools from scratch, e.g., SD system for ATC. We noted that UWB-ATCC corpus was more challenging for the BERT SD model compared to LDC-ATCC and HAAWAII corpora (6.5\% and 8.1\% JER, respectively). However, this system performed consistently better on all the other test sets, if we compare the model fine-tuned on UWB-ATCC versus the ones on LDC-ATCC and HAAWAII. We believe that the transferability to new domain of UWB-ATCC corpus is higher compared to LDC-ATCC and HAAWAII (see `UWB-ATCC' row in Table~\ref{tab:overall_performance} and compare it with LDC-ATCC or HAAWAII). 
\vspace{0.1cm}

\begin{figure}[t]
  \centering
  \includegraphics[width=0.92\linewidth]{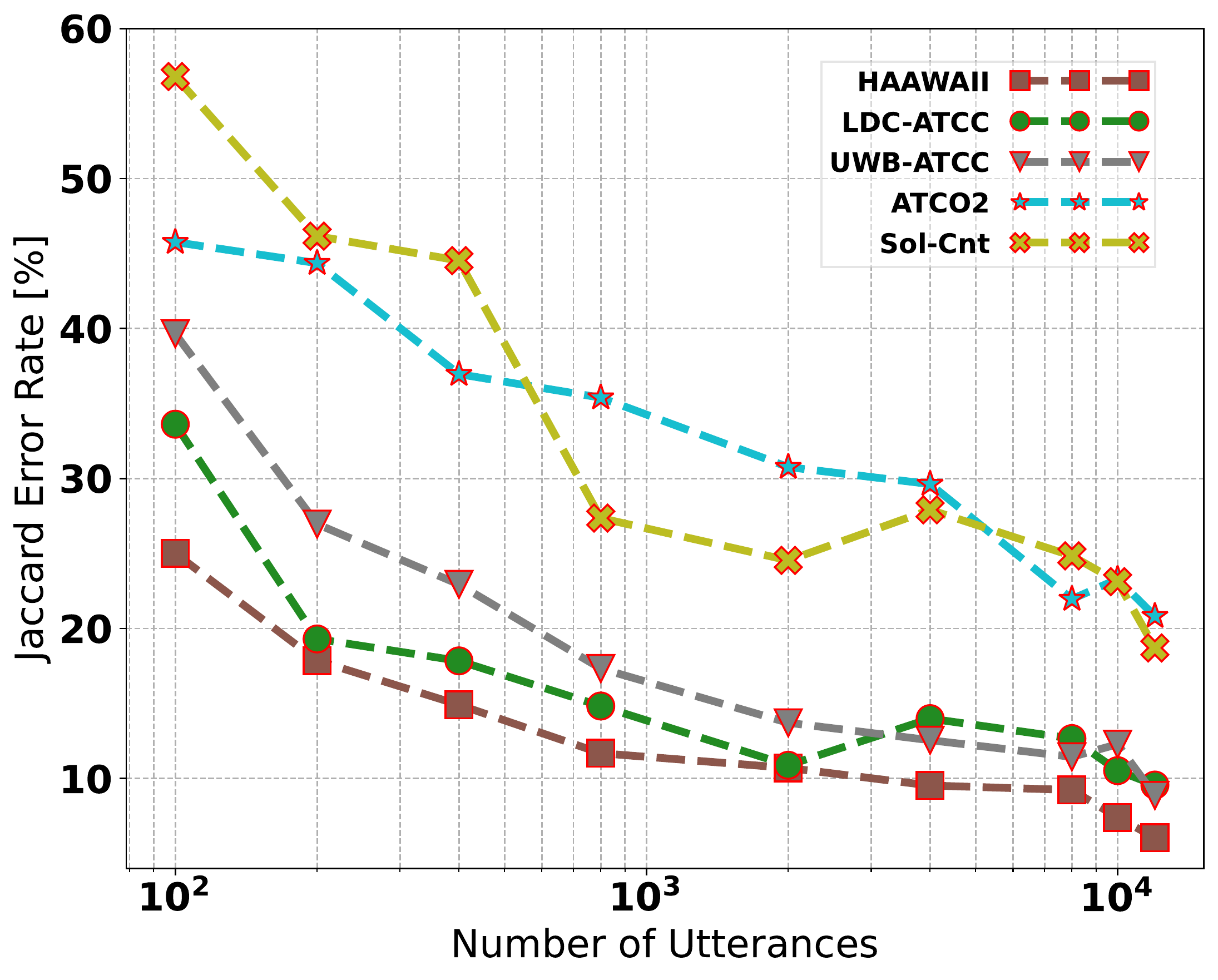} 
  \caption{Jaccard error rates (JER) in percentages (\%) for nine models fine-tuned with increased amount of samples per database. We evaluate models on two configuration. HAAWAII, LDC-ATCC and UWB-ATCC are \textit{in domain experiments}, which means that the train and test splits are from the same database. ATCO2 and SOL-Cnt are \textit{out of domain test sets,} i.e., the train and test data differs. For the two later (blue and yellow dashed lines), we report the results of the model fine-tuned with UWB-ATCC database.}
  \label{fig:incremental_results}
  \vspace{-4mm}
\end{figure}

\noindent \textbf{Does adding more data help?} Here, we evaluate the BERT SD system by performing an ablation where the amount of fine-tuning data is incremental. In total, 9 models per database are studied, as depicted in Figure~\ref{fig:incremental_results} (each data point represents one model). We report token-based JERs which are more aligned to standard SD. 
For the public databases, we obtained 65, 43, and 37\% relative improvement in JERs on LDC, UWB, and ATCO2, respectively, by scaling up the fine-tuning data from 100 to 2000 samples. This number goes up to more than 50\% relative JERs improvement if we use 10k samples (69\% relative improvement for LDC).
We note the same behavior on the private databases. At least 50\% relative improvement is seen by scaling up the data from 100 samples $\rightarrow$ 2000 samples, on both, HAAWAII and SOL-Cnt experiments.
To our surprise, UWB-ATCC models transfer particularly well on the two out-of-domain test sets (i.e., ATCO2 and SOL-Cnt). This gives insights that our approach works well on both, public and private databases. We believe this is an acceptable starting point for the future research on text-based SRD and SCD (not only aligned to ATC).
\vspace{0.1cm}

\noindent \textbf{Robustness of BERT speaker diarization on ASR transcripts:} we evaluated the BERT SD system on \textit{SOL-Cnt} and \textit{UWB-ATCC} test sets, which contain utterances with more than one speaker (\textit{mixed} subset). The BERT SD system is fed with the 1-best transcript obtained from our in-domain hybrid-based ASR system~\cite{nigmatulina2022two}. Table~\ref{tab:der_res} highlights the main results for the BERT SD model, an additional line for \textit{`ASR output'}. In the single-speaker case (either ATCO or pilot), the degradation (ASR transcripts instead of ground truth text) in SD from the BERT SD was no more than 1\% absolute JER and DER (worse, Pilot subset 2.4~$\rightarrow$~3.7\% DER reduction in \textit{SOL-Cnt} set). In the \textbf{MIXED} case, the degradation varied 0.1\% JER and 0.6\% DER absolute in \textit{SOL-Cnt} set, and 3.5\% JER and 1.1\% DER absolute in \textit{UWB-ATCC} set. This behavior is mainly due to the noisy labels produced by the ASR system (see~\cite{zuluaga2020callsign}), i.e., 13\% and 14\% WER on \textit{SOL-Cnt} and \textit{UWB-ATCC} test sets.
\vspace{0.05cm}

\noindent \textbf{Breaking the paradigm, acoustic or text-based speaker diarization?} On challenging tasks such as ATC, where the rate of speech is high and contains mainly close-talk recordings, the standard acoustic-based SD systems are prone to fail and merge two or more segments together. An example is \textit{SOL-Cnt} database (see Table~\ref{tab:test_sets}) where $\sim$38\% of the test set contains more than one speaker or/and segment per utterance (i.e., \textit{`Mixed'}). We compare acoustic-based and BERT SD on private (\textit{SOL-Cnt}) and public (\textit{UWB-ATCC}) test sets. Similar to \textit{SOL-Cnt}, \textit{UWB-ATCC} set contains more than one speaker per utterance. We list the results in Table~\ref{tab:der_res}. In order to contrast both approaches, we compute the JER on the extracted segments, not on the text-level tokens (as done before). Both systems use the same SAD for segmentation. 
The acoustic-based SD, uses the Hungarian algorithm~\cite{jonker1986improving} for assigning the system clusters to the reference speakers. As a result, it evaluates SCD and clustering without identifying the speaker roles. For estimating the DER, we align the text with audio data and prepare the labeled segments from it. Using this alignment, the output of the BERT SD system is comparable to the acoustic-based diarization system. For computing the scores in all systems, the collar of 150 msec was considered. We found out that in noisy conditions, acoustic-based SD mistakenly oversplits the segments with one speaker (either ATCO or pilot). However, the BERT SD seems to be very robust on these segments (3.0/3.7\% $\rightarrow$ 5.8/7.8\% DER for ATCO/pilot of \textit{SOL-Cnt} test set). Even in the mixed scenario of this set, the BERT SD system (9.5\% DER) extended with data augmentation outperformed the acoustic-based model (10.3\% DER) by 7.7\%, relatively. On a cleaner set with shorter segments, VBx system shows the best performance on the segments with one speaker. However, in the mixed segments, the BERT SD system outperformed the VBx by a marginal improvement.

\section{Conclusion}

In this work, we demonstrated that acoustic-based tasks such as speaker diarization can be enhanced or even replaced by natural language processing techniques. Even including challenging tasks such as SD for ATC communications. 
Our results, obtained on examples where SAD failed, validated this hypothesis, as presented in Table~\ref{tab:overall_performance} and Table~\ref{tab:der_res}. 
Additionally, we developed a simple and flexible data augmentation pipeline for ATC text data. To the authors' knowledge, this is the first time that a BERT-based SD could fully replace an acoustic-based SD in the field of ATC. 
We evaluated our approach on public and private datasets in the ATC domain. Our BERT SD model reached up to 10\% and 20\% token-based JER in public and private ATC databases. We compared our model with the well-known acoustic-based SD system (VBx). On the noisy sets, VBx oversplits the segments with one speaker, however, the BERT SD system shows robust performance on these segments. In addition, BERT SD model outperforms VBx by a large margin in segments with more than one speaker (\textit{MIXED}). Finally, we also performed an ablation of the amount of data samples versus performance. 


\section{Acknowledgments}

The work was supported by SESAR Joint Undertaking under Grant Agreement No. 884287 - HAAWAII (\textbf{H}ighly \textbf{A}utomated \textbf{A}ir traffic controller \textbf{W}orkstations with \textbf{A}rtificial \textbf{I}ntelligence \textbf{I}ntegration). The work was also partially supported by CleanSky Joint Undertaking under Grant Agreement No. 864702—ATCO2 (Automatic collection and processing of voice data from air-traffic communications).

\bibliographystyle{IEEEbib}
\bibliography{mybib}

\end{document}